\PassOptionsToPackage{table}{xcolor}
\documentclass[sigconf]{acmart}
\renewcommand\footnotetextcopyrightpermission[1]{}
\usepackage[table]{xcolor}

\newcommand{\systemname}{Deriva-ML}

\newcommand{\Caseproject}{EyeAI}

\AtBeginDocument{%
  }

\acmConference[CHI EA '26]{Extended Abstracts of the 2026 CHI Conference on Human Factors in Computing Systems}{April 13--17, 2026}{Barcelona, Spain}
\acmBooktitle{Extended Abstracts of the 2026 CHI Conference on Human Factors in Computing Systems (CHI EA '26), April 13--17, 2026, Barcelona, Spain}

\begin{document}

\title{Reproducibility Beyond Artifacts: Interactional Support for Collaborative Machine Learning}

\author{Zhiwei Li}
\email{lizhiwei@usc.edu}
\orcid{0009-0003-2848-7711}
\affiliation{%
  \institution{University of Southern California}
  \city{Los Angeles}
  \state{California}
  \country{USA}
}

\author{Carl Kesselman}
\email{carl@isi.edu}
\orcid{0000-0003-0917-1562}
\affiliation{%
  \institution{Information Sciences Institute}
  \city{Marina del Rey}
  \state{California}
  \country{USA}
}

\renewcommand{\shortauthors}{Li and Kesselman}

\begin{abstract}
    Machine learning (ML) reproducibility is often framed as a problem of incomplete artifact recording. This framing leads to systems that prioritize capturing datasets, code, configurations, and execution environments. However, in collaborative and interdisciplinary ML projects, reproducibility failures often arise not only from missing artifacts but from difficulties in interpreting prior work, aligning evolving components, and reconstructing experimental intent over time. Drawing on a 19-month deployment of a data-centric ML management system in a clinical research project, we identify recurring interactional breakdowns that persist despite comprehensive structural traceability. Based on these findings, we propose a two-layer socio-technical ML management system combining lifecycle-aware artifact infrastructure with an interactional layer designed to mediate coordination, explanation, and shared understanding. We discuss how an AI-mediated semantic interface reframes reproducibility as an ongoing socio-technical accomplishment rather than a static property of recorded traces, and outline implications for human-centered ML infrastructure design.
\end{abstract}




\maketitle
\begin{center}
\small
To appear in the Extended Abstracts of the 2026 CHI Conference on Human Factors in Computing Systems (CHI EA '26).
\end{center}

\section{Introduction}
Machine learning (ML) reproducibility remains a persistent challenge, with many ML results difficult to reproduce or meaningfully reuse even when code and data are shared ~\cite{kapoor2023leakage, semmelrock2025reproducibility}. As projects grow more complex and involve more collaborators, reproducibility problems increasingly come from how teams work rather than isolated technical bugs, particularly in high-stakes domains like healthcare, where reproducibility affects trust and clinical validity ~\cite{mcdermott2021reproducibility, beam2020challenges}.

Most reproducibility supporting tools focus on structural support, by which we mean the systematic recording and tracing of ML artifacts such as datasets, code, configurations, execution environments, often coupled with pipeline orchestration ~\cite{kreuzberger2023machine, hernandez2025reproducible}.
Meanwhile, studies of collaborative ML practice emphasize the importance of communication, coordination, and shared understanding across roles, frequently advocating for improved documentation and best practices ~\cite{nahar2022collaboration}. But under deadline pressure and changing goals, teams fall back on informal conversations and tacit knowledge that systems don't capture.

Together, these tensions raise a design challenge for ML infrastructure:
\emph{what human work should ML infrastructure actually support, and how?}
In particular, how can systems help align the technical intent encoded in ML artifacts with the realities of collaborative practice over time?

We address this question through the deployment of \systemname{}, a data-centric and lifecycle-aware ML management system ~\cite{11181473}, within a real interdisciplinary clinical research project, EyeAI ~\cite{nguyen2025comparison}.
By examining reproducibility in a setting where robust structural ML infrastructure was already in place, we identify recurring breakdowns centered on interactional work, such as alignment drift across evolving artifacts and the loss of context embedded in informal communication ~\cite{semmelrock2025reproducibility, bauer2009designing}.

Motivated by these findings, we 
contribute (1) an empirical characterization of interactional reproducibility breakdowns in collaborative ML, and (2) a design framework for AI-mediated interactional infrastructure.
The interactional layer adds conversational support that helps teams navigate the system, understand past work, and preserve informal context. 
Rather than automating ML decisions, this approach scaffolds coordination, interpretation, and shared understanding, and presents a domain-agnostic design space in which ML systems proactively support reproducibility as an ongoing socio-technical accomplishment.

\section{Limits of Artifact-Centric ML Management in~Practice}
\subsection{Structural ML Management in Practice}
To ground our system design exploration, we deployed Deriva-ML within an ongoing interdisciplinary clinical ML project, \Caseproject{}. We treat this 19-month deployment as a reproducibility stress test: rather than examining ML work without tooling, we investigate challenges that persist when structural ML management infrastructure is already in place.

\Caseproject{} reflects common collaborative ML conditions: heterogeneous clinical datasets, evolving objectives, multiple modeling strategies, and coordination across clinicians, ML engineers, and data managers. Over the 19 months, the project produced 80+ datasets, 130 executions, and 3 model types, all versioned and traceable in the system.

Both authors were embedded in the project throughout, participating in regular meetings, managing data and infrastructure, and supporting experiment tracking and reproducibility. This sustained engagement provided visibility into how collaborators interpreted, reused, and struggled with prior work over time.

\subsection{Recurring Breakdowns Beyond Artifacts}

\begin{table*}[t]
\centering
\small
\rowcolors{2}{gray!8}{white}
\begin{tabular}{p{3.2cm} p{4.4cm} p{4.4cm} p{2.8cm}}
\toprule
\textbf{Breakdown pattern} & \textbf{What was missing/insufficient} & \textbf{Consequence in practice} & \textbf{Roles involved} \\
\midrule
Incomplete experiment configurations &
Parameter choices, config files, and ad hoc script changes during experimentation &
Couldn't reproduce runs without asking the original developer & 
ML engineers \\

Environment dependencies &
Keeping data, code, and runtime aligned across framework and dependency changes &
Reproducing results required manual environment reconstruction or retraining &
ML engineers, data managers \\

Dataset version context not fully captured &
Changes in inclusion criteria, preprocessing decisions, and feature engineering &
Models on "similar" datasets couldn't be reproduced one-to-one &
Clinicians, ML engineers, data managers \\

Lost rationale over time &
Why experiments were run, how runs related, why results mattered &
Handoffs needed meetings with original developers; hard to reuse prior work independently &
ML engineers, data managers, clinicians \\

Provenance gaps caused by offline and exploratory work &
Intermediate experiments, failed attempts, exploratory iterations done outside the system &
Reuse required reverse engineering and personal memory &
ML engineers \\

\bottomrule
\end{tabular}
\caption{Recurring interactional breakdowns observed when deploying \systemname{} in a clinical project.}
\label{tab:breakdowns}
\end{table*}

The structured system enabled many experiments to be rerun, compared, and reused successfully. However, during the deployment, we repeatedly encountered situations in which collaborators were unable to interpret or reuse prior experiments solely from recorded artifacts in the system. These incidents surfaced during experiment handoffs, troubleshooting sessions, configuration updating, and retrospective reviews, even when datasets, code, and execution metadata were fully traceable.

Although artifacts were present, collaborators often required additional explanation to reconstruct experimental intent or align evolving components. We documented recurring instances during meetings and troubleshooting exchanges and later revisited them, iteratively grouping them into breakdown categories through collaborative reflection between the authors. As embedded infrastructure designers and participants, our analysis reflects this situated perspective and presents analytically derived patterns rather than statistically representative findings. Table~\ref{tab:breakdowns} summarizes the resulting analytical patterns that reveal limits of artifact-centric support.







These breakdowns cluster into three patterns. First, alignment drift: as datasets, scripts, and environments evolved separately, understanding how changes related across iterations required knowledge beyond what was structurally encoded. Second, exploratory work outside the system: under deadline pressure or during early experimentation, developers sometimes worked outside Deriva-ML, requiring later reconstruction through memory or discussion. Third, loss of shared context: as experiments accumulated, understanding their rationale increasingly depended on the original developers, making handoffs reliant on synchronous explanation.

Across all three, reproducibility failed not because artifacts were missing, but because the reasoning and context surrounding them were not readily accessible later. This suggests that sustaining reproducibility requires supporting ongoing explanation, coordination, and shared understanding, not just artifact tracking.

\subsection{From Interactional Breakdowns to Design~Goals}

The patterns above reveal that reproducibility problems in collaborative ML rarely stem from missing artifacts. 
Even with comprehensive infrastructure, difficulties arise when collaborators revisit, extend, or hand off prior work, requiring explanation of decisions, relationships across iterations, and experimental intent. The challenge is not incomplete records, but a mismatch between how systems encode work and how people reason about it.

These observations suggest that reproducibility in collaborative ML work is not a static property of artifact completeness, but an ongoing socio-technical accomplishment. 
While structural ML management systems provide essential foundations for traceability and repeatable execution, they do not actively support the human work through which reproducibility is sustained over time. This includes sensemaking across iterations, coordination during exploration, and the articulation of informal decisions that shape experimental trajectories.

The breakdowns point to three design goals for ML management infrastructure:

\textbf{Help maintain alignment as work evolves.} Address recurring
misalignments across configurations, dependencies, and dataset versions (Table 1, rows 1–3) by helping collaborators understand how changes relate, which dataset versions work with specific code, why parameters shifted, and how experimental goals evolved.

\textbf{Support exploration within system boundaries.} Reduce the need for informal or offline experimentation (rows 1 and 5) by helping people understand system expectations while experimenting, so exploratory work remains connected to structured tracking.

\textbf{Preserve informal context as shared artifacts.} Mitigate the loss of rationale and contextual decision-making (rows 3–5) by capturing discussions, assumptions, and negotiated trade-offs as accessible project knowledge, especially during handoffs.

Meeting these goals requires infrastructure that actively bridges the gap between human goals and system knowledge. Instead of treating coordination as something that happens outside the system, it should be supported directly alongside the artifact tracking.
This shift requires designing interfaces that make encoded system knowledge conversationally accessible during ongoing work, rather than leaving interpretation to ad hoc meetings or personal memory.

\section{A Two-Layer Socio-Technical ML Management System}
\begin{figure*}[t]
  \includegraphics[width=\textwidth]{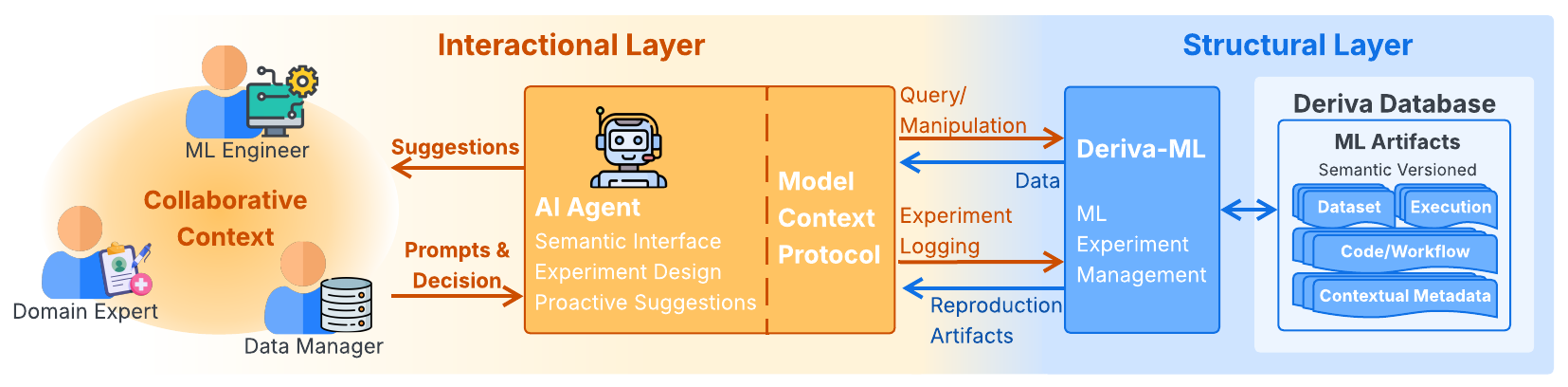}
  \caption{A two-layer socio-technical ML management system. The structural layer captures ML artifacts and their evolution. The interactional layer mediates collaboration through an AI agent that provides practice guidance, experiment suggestions, and decision support, communicating with the structural layer via the Model Context Protocol (MCP).}
  \Description{Diagram of a two-layer machine learning management system. The structural layer represents datasets, code, executions, and metadata with explicit relationships. The interactional layer sits above this structure and supports human collaboration, interpretation, and communication around evolving ML artifacts.}
  \label{fig:architecture}
\end{figure*}

We present a two-layer socio-technical ML management system (Figure~\ref{fig:architecture}) that integrates a \emph{structural layer} for managing evolving artifacts with an \emph{interactional layer} that mediates collaboration. \footnote{A demonstration video showing example interaction workflows is included in the supplementary materials.}

\subsection{Structural Layer: Data-Centric, Lifecycle-Aware ML Infrastructure}
The structural layer, shown on the right side of Figure~\ref{fig:architecture}, is implemented using \systemname{}, which encodes ML development as a collection of formally represented artifacts that evolve over time. The system prioritizes explicit versioning, provenance, and lifecycle alignment over informal or ad hoc representations~\cite{11181473}. It provides a robust Python-based API, along with a rich and highly configurable browser-based user interface~\cite{bugacov2017experiences}.

For clarity, Figure~\ref{fig:architecture} abstracts this structure into four conceptual categories: data, code and workflows, executions, and contextual metadata, which are semantically versioned~\cite{semver2000_prestonwerner} and linked through explicit provenance relationships. This structure supports core reproducibility tasks such as tracing experiment lineage, comparing runs across iterations, and rerunning workflows under controlled conditions. In practice, this information is accessed through interfaces such as the Chaise UI ~\cite{10.1145/3400903.3400908}, which expose artifact relationships but do not capture the informal reasoning surrounding their use.

As shown in Section~2, this form of structural traceability is essential but insufficient for sustaining reproducibility in collaborative practice. While the structural layer records what was executed and how artifacts are related, it does not support the explanation, negotiation, or sensemaking through which collaborators interpret prior work and align evolving experiments with shared intent. These limitations motivate the interactional layer described next.

\subsection{Interactional Layer: AI-Mediated Support for Human-Centered ML Practice}

To address the interactional gaps identified in Section~2, we introduce an interactional layer that provides context-driven scaffolding on top of the structural infrastructure. As shown on the left side of Figure~\ref{fig:architecture}, this layer centers on an AI agent that acts as a semantic interface between collaborators and the ML management system, helping align human goals with encoded system expectations. 
The agent is implemented using a large language model (Claude) with access to \systemname{} through Model Context Protocol (MCP) tool calls: typed, schema-aware functions that expose artifact queries, provenance traversal, and execution comparisons against the live system state.

A central design principle of the interactional layer is that the agent is grounded in the semantics of the underlying ML management system. Because the agent has access to the structured representation of datasets, workflows, executions, configurations, and their relationships, it can support interaction at a level that reflects how the system itself understands ML work. 

A crucial aspect of the interactional layer is that it embeds reproducibility within users’ primary objective of developing effective ML models, rather than treating it as a separate task. By supporting rapid and exploratory model development directly within the system, the interface aligns everyday practice with structured tracking. In this way, reproducibility becomes an integrated outcome of normal use, reducing the likelihood that users bypass infrastructure requirements under time pressure.

The interactional layer also supports experiment configurations as a collaborative and interpretive activity. For example, a collaborator asking which executions used a given dataset version can receive a provenance-grounded answer linking dataset history, parameter choices, and downstream runs, reconstructing the rationale that would otherwise require tracking down the original developer. Drawing on its knowledge of valid configurations, dependencies, and lifecycle constraints encoded in \systemname{}, the agent helps practitioners set up experiment configurations, identify misalignments, and clarify how configuration choices relate to prior work.

Finally, the interactional layer treats human interactions as traceable artifacts rather than ephemeral exchanges. Interactions with the agent generate dialog transcripts that capture rationales, assumptions, and negotiated decisions that would otherwise remain ephemeral. These records accumulate as shared contexts linked to evolving artifacts that support reflection, handover, and reuse over time. In this way, the agent helps to preserve unstructured interactional knowledge as part of the ML development record, addressing the loss of shared understanding observed in long-running collaborative projects.

The interactional layer is currently implemented as a prototype and has not yet been evaluated through longitudinal deployment. It has been tested on small-scale pipeline tasks including data ingestion, dataset creation, feature generation, and ML fine-tuning configuration, but not formally validated at scale. Its role in this paper is to demonstrate how AI agents can function as interactional mediation within the ML management infrastructure, making encoded system knowledge accessible and actionable for human practice rather than merely automating technical tasks.


\section{Discussion}

\subsection{Reproducibility as Interactional Work}

Prior work on ML  reproducibility has largely emphasized the recording and sharing of technical artifacts, including datasets, code, configurations, workflows, and execution environments ~\cite{hernandez2025reproducible, semmelrock2025reproducibility}. This focus underlies many ML management and MLOps tools, as well as practices such as containerization and standardized documentation, which aim to support repeatable execution and post hoc auditability ~\cite{hernandez2025reproducible}. 
Yet even when such infrastructure is in place, collaboration challenges persist: studies of ML teams document fragmented documentation, informal communication dependencies, and coordination failures across roles~\cite{nahar2022collaboration, piorkowski2021ai, amershi2019software}. 
Research on MLOps further suggests that while tooling can improve certain aspects of coordination, it may also introduce new burdens related to integration, ownership, and maintenance ~\cite{sothilingam2022using}.

Our findings connect these strands of work by showing how collaboration challenges surface specifically as reproducibility breakdowns, even when robust artifact management infrastructure is in place. The breakdowns we observed did not stem from missing datasets or code, but from difficulties in aligning evolving artifacts, interpreting recorded work in light of prior decisions, and reconstructing context during reuse and handoff. From this perspective, reproducibility is not a static property of recorded traces, but an ongoing socio-technical accomplishment that depends on interactional work such as explanation, coordination, and shared understanding. This reframing suggests that ML infrastructure must support not only artifact traceability but also the human practices through which reproducibility is sustained over time.

\subsection{AI Agents as Interactional Support in ML Infrastructure}

Recent advances in AI-assisted development environments demonstrate the value of embedding intelligent, context-aware assistance directly into everyday work practices. AI-facilitated IDEs and emerging AI-enabled MLOps tools increasingly support tasks such as workflow configuration, debugging, and experiment management~\cite{peng2023impact, sergeyuk2026human}. These developments show that AI can operate effectively within complex technical systems without replacing human judgment.

Our work explores a different role for AI in ML infrastructure: supporting the interactional work that enables reproducibility and collaboration. 
While the agent assists with ML experiment configuration, its primary role is to serve as a semantic interface that helps practitioners understand system expectations, navigate complex experiment histories, and preserve informal decisions and rationales as shared context.
This addresses the breakdowns from our deployment, where system knowledge existed but was difficult to access or act upon in practice.

The key is tight coupling between the conversational interface, human collaborators, and the underlying infrastructure. 
Because the system accesses the full project context (domain questions, artifact database, codebase), it can assist in sensemaking and decision-making.    
This creates a feedback loop in which conversations produce richer descriptions and rationales around artifacts, which in turn enable the system to provide better guidance when people interpret results, spot misalignments, or decide what to try next.

We argue that this feedback loop strengthens both human coordination and system traceability. 
Rather than treating human expertise, AI assistance, and infrastructure as separate components, the design enables them to mutually reinforce one another. 
In doing so, it points toward a shift from artifact-centric ML systems toward socio-technical systems that support interpretation and coordination over time.


\section{Limitations and Future Work}
This work is based on one interdisciplinary ML project, which limits how broadly the breakdown patterns apply. We see these patterns as diagnostic rather than project-specific, but studying teams in other domains and configurations would show how interactional challenges vary across settings.
The interactional layer is an early prototype tested on small examples. 
Longitudinal deployment with real teams will show how conversational support 
affects coordination, interpretation, and handoffs in practice.
Finally, our embedded role as infrastructure developers provided sustained access to collaborative ML work but may have shaped our interpretations. Studying similar systems across multiple teams will help validate and refine the proposed design space.

\bibliographystyle{ACM-Reference-Format}
\bibliography{sample-base}



\end{document}